\documentstyle[11pt,paspconf]{article}

\def\etal{{\it et al.}}

\begin{document}

%
%

\title{Discovery of a Galaxy Responsible for a DLA System at $z=3.15$ and a
	Near-Infrared Search for Primeval Galaxies}

\author{Michael A. Pahre and S. G. Djorgovski
}
\affil{Palomar Observatory, Caltech, MS 105--24, Pasadena, CA   91125}

\author{Jill Bechtold}
\affil{Steward Observatory, The University of Arizona, Tucson, AZ  85721}

\author{Richard Elston}
\affil{Cerro Tololo Inter-American Observatory, Casilla 603, La Serena,
	Chile}


%
%

\begin{abstract}

We report the detection, both in line and continuum emission, of a galaxy 
responsible for a damped Ly$\alpha$ absorption (DLA) system at $z=3.15$ 
projected $2.3''$ from QSO~2233+131.  The star formation rate 
implied by both the Ly$\alpha$ and restframe far-UV continuum is 
$\sim 7$~M$_\odot$~yr$^{-1}$.  The galaxy also shows no sign of an active 
nucleus and has blue colors, both consistent with it being observed early in 
its star formation history.  The Ly$\alpha$ emission is offset 
$\sim 200$~km~s$^{-1}$ from the DLA velocity, which could be explained as 
the rotational signature of a disk galaxy.

We also report on the progress of an ongoing near-infrared narrow-band search 
with the Keck telescope for line emission from a population of primeval 
galaxies at high redshift.  We have found several promising candidates
which are being followed up with optical spectroscopy.  Our survey is
beginning to place constraints on a possible population of slightly dusty 
primeval galaxies at high redshift.

\end{abstract}



%
%




%
%

\section{Discovery of the Galaxy Responsible for DLA~2233+131}


Damped Lyman-alpha absorber (DLA) systems, detected along lines of sight to 
unrelated high-redshift quasars, have been proposed as possible progenitors 
of normal disk galaxies today.  They appear to contain a substantial
fraction of the baryons known to exist in normal galaxies today (Lanzetta,
Wolfe, \& Turnshek 1995).  Several searches for the galaxies responsible
for various DLA systems have resulted in a few detections of emission--line
galaxies (Lowenthal \etal\ 1991; Macchetto \etal\ 1993; M\o ller \& Warren 
1993; Francis \etal\ 1996), but all appear either to be AGN or 
are at a similar redshift as the QSO, suggesting that the source of their 
ionization may not be a result of star formation.

We report on the detection of an object,
designated DLA 2233+131, responsible for a previously known DLA system at
$z_{abs} = 3.150$ (Lu \etal\ 1993) in the spectrum 
of QSO~2233+131 ($z_{QSO} = 3.295$; Crampton, Schade, \& Cowley 1985). 
It was found serendipitously during observations of a candidate for a 
different DLA system in the same field.  The
object was also selected independently as a DLA candidate on the basis of 
its broad-band colors by Steidel, Pettini, \& Hamilton (1995), which was 
unknown to us at the time.  We obtained direct images of the field in the 
$V_{\rm C}$ and $R_{\rm C}$ bands, and long-slit spectra using low and 
moderate resolution gratings.
Details can be found in Djorgovski \etal\ (1996).
The discovery spectrum of the DLA is shown in Figure 1; the QSO spectrum
is plotted for comparison.

\begin{figure}
\plotfiddle{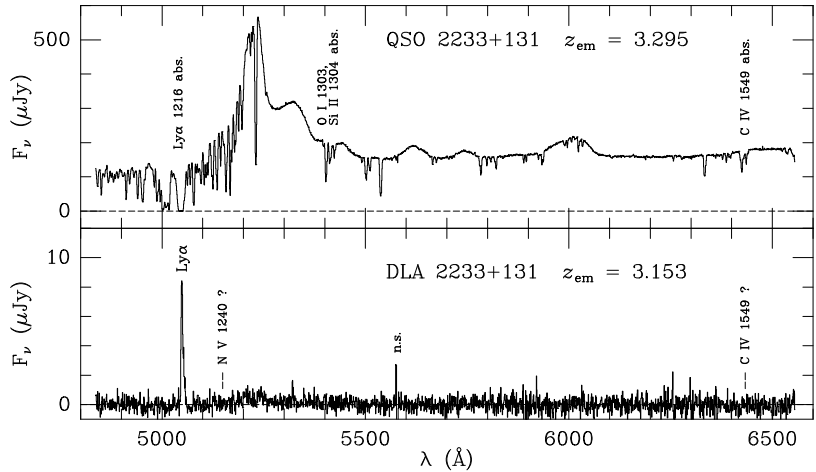}{2.5in}{-90}{50}{50}{-200}{250}
\caption{The spectra of QSO~2233+131 [top] and DLA~2233+131 [bottom].
The DLA is seen in emission $2.3''$ projected from the QSO.  The DLA shows
emission from Ly$\alpha$ at $z=3.1530$, and no signs of higher-ionization 
lines that would be expected for an active nucleus.
\label{fig1}}
\end{figure}

We find the DLA in emission at a location offset $2.3''$ 
($17.2$~kpc for $H_0=75$ and $\Omega_0=0.2$) from the QSO line--of--sight.  
There are no signs of any higher--ionization emission lines from the DLA, 
suggesting that this object does not harbor an AGN.  The star formation rate 
implied by the Ly$\alpha$ line is $7.5 {\rm~M}_\odot {\rm~yr}^{-1}$, 
and that by the restframe far--UV continuum is 
$6.4 {\rm~M}_\odot {\rm~yr}^{-1}$, suggesting
that there is very little dust present in this galaxy.  The Ly$\alpha$
luminosity for this galaxy is the highest for non--AGN at these redshifts
in the list of Steidel \etal\ (1996).  This galaxy's luminosity is similar
to that of a $L^\ast$ galaxy today.

The emission line appears offset by $\sim 200$~km~s$^{-1}$ from the
damped absorption line, which could be explained by a rotating disk
galaxy.  DLA systems represent a significant population of high--redshift
objects which have been proposed as the progenitors of normal disk
galaxies (Wolfe 1993).  Our observations for DLA~2233+131 are fully--consistent
with that hypothesis.  HST images are planned for this source (C. Steidel,
private communication), which may be capable of determining if the source
shows a disk--like morphology or one more similar to that of the Steidel \etal\
population (e.g. in Giavalisco \etal\ 1996).

%
%

\section{A Near-Infrared Search for Line Emission from Protogalaxies}


The search for primeval galaxies---a population of the progenitors of 
present--day spheroidal galaxies---is a central goal of modern observational
cosmology.  Recent optical searches for the signs of Ly$\alpha$ emission
from such a population (e.g. Thompson, Djorgovski,
\& Trauger 1995) have been largely unsuccessful.  Those optical
searches for Ly$\alpha$ could be brought into agreement with the present-day
space density of spheroidal systems if the primeval galaxies were, in general,
enshrouded in a modest quantity of dust.  

Similar searches for longer--wavelength emission lines should be much
less-affected by dust obscuration.  For this reason, we are pursuing a
search for the O~II, H$\beta$, [O~III], and H$\alpha$ lines at redshifts
of $2 < z < 5$ using the near-infrared camera on the W. M. Keck Telescope.  
We have deliberately chosen fields centered on 
high redshift QSOs, DLAs, or radio galaxies, and narrow-band filters to match 
an emission line at that known object's redshift.  Results for the first 
four fields were presented by Pahre \& Djorgovski (1995);  to date we have 
surveyed nearly three times the area described in that work, and are 
reaching flux limits up to twice as deep.  The estimated limits imposed by 
our survey are shown in Figure 2.
Comparison is made to several other near-infrared
searches:  Thompson, Djorgovski, \& Beckwith (1994, TDB); Parkes, Collins,
\& Joseph (1994, PCJ); and Mannucci, Beckwith, \& McCaughrean (1994, MBM).
(See Pahre \& Djorgovski 1995 for further details on the comparison.)
In our search we have found a few promising candidates with narrow--band
excess emission of $> 3 \sigma$, and are currently following these up with 
optical spectroscopy.

\begin{figure}
\plotfiddle{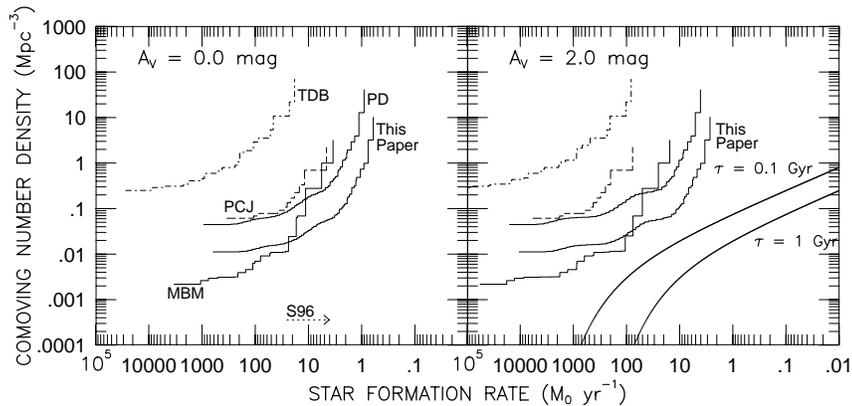}{2in}{0}{50}{50}{-210}{-75}
\caption{
The estimated limits set by near-infrared narrow-band searches
for emission-lines due to primeval galaxies.  Regions of the figure above
and left of the lines are excluded by the observations.  The effects due to
a simple dust screen model with $A_V=2$~mag, which is required to explain the
lack of detection of Ly$\alpha$ for optical searches for primeval galaxies,
is shown in the right panel.  
The population of $z>3$ star-forming galaxies discovered by Steidel \etal\
(1996, S96) is shown at the bottom of the left panel; the effects of resonant
scattering on the Ly$\alpha$ line are shown with an arrow.
Two single--burst models are shown in the right panel and labelled with their 
star-formation timescales.
\label{fig2}}
\end{figure}

As shown in Figure 2, we are beginning to probe the relevant region of
parameter space where we would expect to find a population of primeval
spheroidal galaxies if they were to form during a relatively short burst
($\tau < 0.1$~Gyr) of star formation while obscured by a small
quantity of dust.  We note that the population of star--forming galaxies
discovered by Steidel \etal\ (1996) should be undetected in our survey
because of their low star formation rates and surface density.  On the
other hand, if there exists a population of highly--reddened, star--forming 
galaxies at these redshifts they would be missed by the blue selection
criteria of Steidel \etal\ but should in principle be found by our search 
for the longer--wavelength nebular lines.  Furthermore, the importance of
resonant scattering of the Ly$\alpha$ found by Steidel \etal\ should not
affect the nebular lines used in our near-infrared search.

%
%

\acknowledgments

We would like to thank L. Lu, M. Rauch, C. Steidel, and D. Thompson for
many useful discussions.  This work was supported in part by NSF PYI
award AST-9157412 and the Bressler Foundation.  
We thank the staff of W. M.
Keck Observatory and Palomar Observatory for the expert help during our
observing runs.

%
%

\end{document}